\documentclass[conference]{IEEEtran}
\IEEEoverridecommandlockouts
\usepackage{cite}
\usepackage{amsmath,amssymb,amsfonts}
\usepackage{algorithmic}
\usepackage{graphicx}
\usepackage{float}

\usepackage{algorithmic}
\usepackage[ruled,vlined]{algorithm2e}
\usepackage{cuted, nccmath}
\usepackage{caption}
\usepackage{subcaption}
\usepackage{textcomp}
\usepackage{xcolor}
\usepackage{subfiles}
\def\BibTeX{{\rm B\kern-.05em{\sc i\kern-.025em b}\kern-.08em
    T\kern-.1667em\lower.7ex\hbox{E}\kern-.125emX}}
\begin{document}

\title{Dense Error Map Estimation for MRI-Ultrasound Registration in Brain Tumor Surgery Using Swin UNETR}

\author{
\IEEEauthorblockN{Soorena Salari\textsuperscript{1}, Amirhossein Rasoulian\textsuperscript{1}, Hassan Rivaz\textsuperscript{2}, Yiming Xiao\textsuperscript{1}}
\IEEEauthorblockA{\textsuperscript{1} Department of Computer Science and Software Engineering, Concordia University, Montreal, QC, Canada \\
\textsuperscript{2} Department of Electrical and Computer Engineering, Concordia University, Montreal, QC, Canada \\
soorena.salari@concordia.ca, ah.rasoulian@gmail.com,\\
hassan.rivaz@concordia.ca, yiming.xiao@concordia.ca}
}







\maketitle

\begin{abstract}

Early surgical treatment of brain tumors is crucial in reducing patient mortality rates. However, brain tissue deformation (called brain shift) occurs during the surgery, rendering pre-operative images invalid. As a cost-effective and portable tool, intra-operative ultrasound (iUS) can track brain shift, and accurate MRI-iUS registration techniques can update pre-surgical plans and facilitate the interpretation of iUS. This can boost surgical safety and outcomes by maximizing tumor removal while avoiding eloquent regions. However, manual assessment of MRI-iUS registration results in real-time is difficult and prone to errors due to the 3D nature of the data. Automatic algorithms that can quantify the quality of inter-modal medical image registration outcomes can be highly beneficial. Therefore, we propose a novel deep-learning (DL) based framework with the Swin UNETR to automatically assess 3D-patch-wise dense error maps for MRI-iUS registration in iUS-guided brain tumor resection and show its performance with real clinical data for the first time.

\end{abstract}

\begin{IEEEkeywords}
Deep Learning, Registration, Inter-modal, Error grading
\end{IEEEkeywords}

\section{Introduction}
Removing brain tumors in early stages can significantly decrease the mortality rate of patients. However, brain tissue deformation (called brain shift) can happen during surgery due to several factors like gravity, drug administration, and pressure changes after craniotomy. Although modern MRI approaches can deliver precise anatomical and physiological details for pre-surgical planning, intra-operative MRI is expensive and requires a complicated setup to track brain shift. In contrast, intra-operative ultrasound (iUS) is more popular due to its lower cost, portability, and flexibility in real-time imaging during surgery \cite{rivaz2015near}. Accurate registration techniques between MRI and iUS \cite{xiao2019evaluation} can significantly enhance the value of iUS for updating pre-surgical plans and guiding surgical understanding. However, manual inspection of registration results is tough and prone to error due to the 3D nature of surgical data and time constraints. Thus, algorithms that can detect unreliable inter-modal medical image registration outcomes are invaluable for precision-sensitive neurosurgery.

The automatic quantification of medical image registration quality has become increasingly important in the fields of medical image computing and surgical interventions \cite{bierbrier2022estimating}. With advancements and progress in machine learning (ML) and deep learning (DL) methods in other fields, ML and DL techniques have been utilized to efficiently estimate errors in medical image registration. Early research relied on hand-crafted features \cite{shams2017assessment,saygili2018local,sokooti2019quantitative,sokooti2016accuracy,eppenhof2018error,sokooti2021hierarchical}, while more recent efforts have utilized DL methods to automatically assess intra-contrast/modality applications such as CT \cite{eppenhof2018error,sokooti2021hierarchical} and MRI \cite{fonov2022darq}. However, there has been little exploration of error estimation in inter-contrast/modal registration, which is of paramount importance for surgical applications. Recently, Bierbrier \textit{et al.} \cite{bierbrier2023towards} attempted to address this gap by training 3D convolutional neural networks using simulated iUS from MRI for MRI-iUS registration in ultrasound-guided brain tumor resection. However, their framework only worked well on simulated data, and further improvements are still needed for real clinical datasets. Also, Salari \textit{et al.} \cite{salari2023focalerrornet} proposed a network for mean error estimation of multi-modal image registration, but their framework didn't produce dense error maps.

In this paper, we propose a novel DL-based framework with Swin UNETR \cite{hatamizadeh2021swin,tang2022self} to automatically evaluate patch-wise dense error maps for MRI-iUS registration in ultrasound-guided brain tumor resection and demonstrate excellent performance with real clinical data for the first time.

\begin{figure*}[h]
\centering
\includegraphics[scale=0.45]{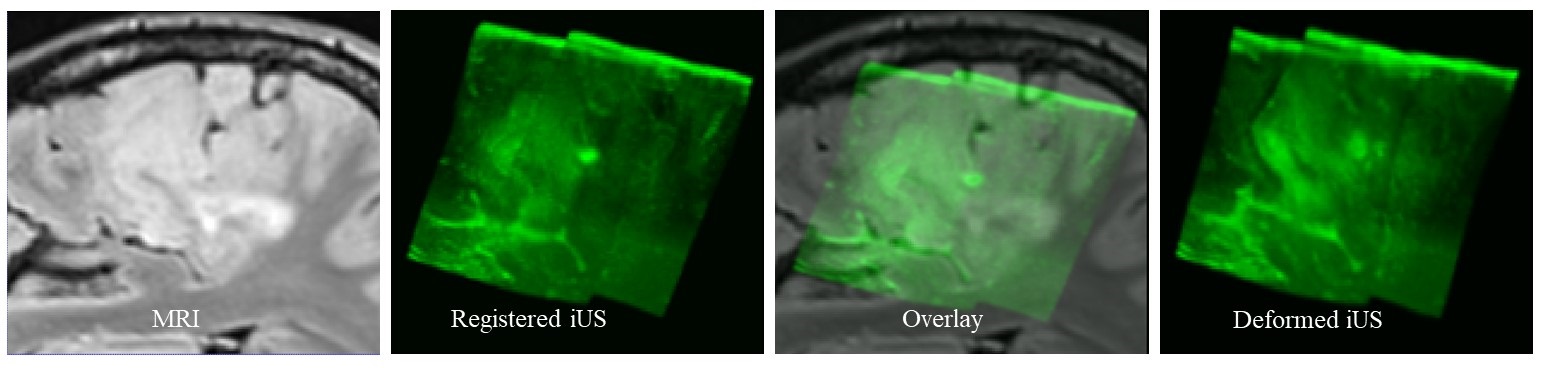}
\caption{An example of an MRI-iUS pair from a patient with registered iUS volume-based B-spline registration and deformed iUS with a mean registration error of 1.4 mm.}
\label{Pair}
\end{figure*}

\section{Experimental setup}
\subsection{Data preprocessing}
To develop and evaluate our methodology, we utilized the RESECT (REtroSpective Evaluation of Cerebral Tumors) dataset \cite{xiao2017re} containing pre-operative MRI and iUS scans from 23 individuals who experienced low-grade glioma resection surgeries. Since modeling iUS scans with tissue resection is challenging, we picked 22 cases with T2FLAIR MRI that better exhibit tumor boundaries and iUS scans taken before surgery. Clinical iUS can deliver more realistic image features and potentially better outcomes in clinical applications than simulated contrasts \cite{eppenhof2018error,bierbrier2023towards}. However, acquiring an accurate brain shift model is impractical, so we generated silver ground truths for image alignment by using homologous landmarks between MRI and iUS to perform landmark-based 3D B-Spline nonlinear registration for all 22 cases. To address the limited field of view (FOV) in iUS, we cropped the T2FLAIR MRI to fit the FOV of iUS.
Also, we resampled the T2FLAIR MRI and iUS to a $0.5 \times 0.5 \times 0.5$ $\mathrm{mm}^3$ resolution. We then added simulated random misalignment to the iUS scans to build and test our DL model. Here, we also utilized 3D B-Spline transformation, which has been previously discussed in similar studies \cite{lotfi2013improving,sokooti2021hierarchical,bierbrier2023towards}, for the purpose of implementing spatial misalignment augmentation. Figure \ref{Pair} displays a pair of MRI and iUS images taken from a patient.

In short, the B-Spline transformation can be represented by a grid of evenly distributed control points and their corresponding parameters, enabling different degrees of nonlinear distortion. The spacing of the control points determines the levels of detail in local deformation fields, while the displacement parameters specify the enlargement of the deformation. In order to have a diverse range of simulated registration errors, we used a random selection process to determine the number of control points and their displacements in each 3D axis. The maximum values for this selection were set at 20 points and 10 mm, respectively. It should be noted that the control point grid is isotropic, and the density is randomly specified per deformation in our case. We deformed each co-registered iUS scan for a total of ten times. After applying misalignment augmentation to the iUS that was formerly co-registered, we chose matching pairs of 3D image patches measuring $64 \times 64\times 64$ voxels from both the iUS volume and the corresponding MRI. To ensure that the patches we obtained from iUS contain helpful information, we concentrated on acquiring patches centered around anatomical landmark locations provided by the RESECT database. This is because iUS has a limited FOV for the brain tissue and may not display anatomical features in some portion of the 3D reconstructed volume. Because the B-spline transformation provides a displacement vector for each voxel in the iUS volume, we simply used the magnitude of the vector as the simulated registration error for that voxel. Ultimately, the image patch pairs and their corresponding registration errors were utilized in the training and validation process of the suggested DL algorithm. 


\subsection{Network architecture}
\begin{figure*}[h]
\centering
\includegraphics[scale=0.3]{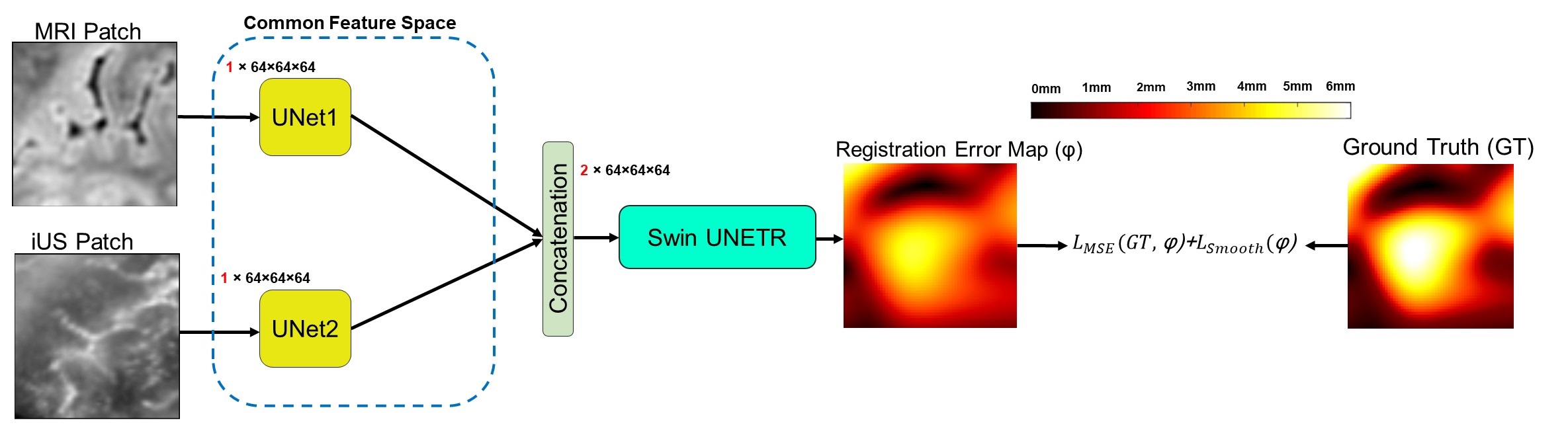}
\caption{An overview of the proposed deep learning framework to obtain dense error map estimation for MRI-iUS registration in ultrasound-guided brain tumor resection.}
\label{Framework}
\end{figure*}

In this paper, we proposed a novel DL-based framework with Swin UNETR \cite{hatamizadeh2021swin} for registration quality control. The overview of the proposed framework is shown in Fig. \ref{Framework}. In this framework, two 3D UNet were employed to separately encode the features of MRI and iUS and bring them to the common feature space. After feature transformation from the 3D UNets, we concatenated the extracted features and fed them to a Swin UNETR model to produce the registration error maps. The reason for choosing Swin UNETR is that it has shown to be highly effective in various downstream computer vision tasks due to its ability to capture both local and global features in complex visual data through the self-attention mechanism of the Swin Transformer \cite{hatamizadeh2021swin,tang2022self}.

\subsection{Evaluation and Implementation Details}
In total, we obtained 3380 patches. The image patches were split subject-wise (22 subjects) into training, validation, and test sets (60\%:20\%:20\%). The incorporated Swin UNETR was pre-trained on 3D CT images (The Beyond the Cranial Vault (BTCV) abdomen challenge and Medical Segmentation Decathlon (MSD) datasets), and we used the pre-trained weights provided by Tang \textit{et al.} \cite{tang2022self}. Then we used this pre-trained model in our framework and trained the framework end-end with our samples. The hyperparameters for network training are specified as defined in Table \ref{Hyp}.

\begin{table}[h]
\centering

\caption{Hyperparameters for the model training.}
\begin{tabular}{ccc p{3cm} p{3.9cm} p{2.5cm} p{1.1cm} p{1.5cm} p{2.5cm}}
\hline

Hyperparameter & Amount \\

\hline
\hline

Image Size &  $64 \times 64\times 64$   \\
\hline

Batch Size &  8   \\
\hline

Epochs &  200   \\
\hline

Optimizer &  AdamW   \\
\hline

Learning rate &  0.0001   \\
\hline

\end{tabular}
\label{Hyp}
\end{table}

\noindent We chose the loss function based on the mean squared error (MSE) that quantifies the predicted registration error map and ground truths. Inspired by the work of Balakrishnan \textit{et al.} \cite{balakrishnan2019voxelmorph}, we have added a regularization term to our loss function, which is the norm of the gradients of the predicted registration error map. This regularization helps enforce the smoothness of the error map. The total loss function can be written as:

\begin{equation}
\mathcal{L}_{\text {sim }}(f,\phi)+\lambda \mathcal{L}_{\text {smooth }}(\phi)
\end{equation}

\noindent where $\phi$ represents the predicted registration error map, $f$ is the ground truth, function $\mathcal{L}_{\text {sim }}(\cdot, \cdot)$ determines mean squared error (MSE) between its two inputs, $\mathcal{L}_{\text {smooth }}(\cdot)$ imposes regularization, and $\lambda$ is the regularization trade-off parameter. In our simulations, we considered $\lambda$ as 0.01.

\section{Results}
To assess the performance of 3D patch-wise registration error estimation, we used two main metrics. First, the accuracy of the 3D patch-wise registration error map was evaluated against the generated ground truths using mean absolute errors (MAEs). Second, we measured the average runtime for an estimation to assess the efficiency of the proposed method. The case-by-case results are shown in Table \ref{SubjectResult} with the respective patient IDs. As indicated in Table \ref{Result}, our method resulted in a low MAE of 0.5mm±0.26mm on test patches, and the average runtime for each estimation was 1.77 s. Our proposed method possessed high clinical potential in neurosurgeries.

\begin{table}[htbp]
\centering

\caption{Case-by-case results of the proposed framework.}
\begin{tabular}{ccc p{4cm}p{4.5cm}p{4.5cm}p{4.5cm}p{4.5cm}p{4.5cm}}
\hline
 Patient ID  & MAE (mm) \\

\hline
\hline
1 &  0.27$\pm$0.03 \\
\hline

2 &  0.45$\pm$0.07 \\
\hline

3 &  0.62$\pm$0.09 \\
\hline

4 &  0.62$\pm$0.12\\
\hline

5 &  0.59$\pm$0.23 \\
\hline

6 &  0.71$\pm$0.05\\
\hline

7 &  0.65$\pm$0.10\\
\hline

8 &  0.64$\pm$0.08 \\
\hline

12 &  0.76$\pm$0.09 \\
\hline

13 &  0.27$\pm$0.04\\
\hline

14 &  0.78$\pm$0.03\\
\hline

15 &  0.76$\pm$0.07\\
\hline

16 & 0.34$\pm$0.14\\
\hline

17 &  0.63$\pm$0.18\\
\hline

18 &  0.33$\pm$0.11\\
\hline

19 &  0.50$\pm$0.06\\
\hline

21 &  0.76$\pm$0.11\\
\hline

23 &  0.39$\pm$0.21\\
\hline

24 &  0.91$\pm$0.11\\
\hline

25 &  0.43$\pm$0.05\\
\hline

26 &  0.19$\pm$0.06\\
\hline

27 &  0.81$\pm$0.21\\
\hline
\hline

\textbf{Mean} &  \textbf{0.56}$\pm$\textbf{0.10}\\

\hline

\end{tabular}
\label{SubjectResult}
\end{table}

\begin{table}[h]
\centering

\caption{Results of the proposed framework.}
\begin{tabular}{ccc p{3cm} p{3.9cm} p{2.5cm} p{1.1cm} p{1.5cm} p{2.5cm}}
\hline

MAE (mm) & Average run time (s) \\

\hline
\hline

0.5$\pm$0.26 & 1.77 \\
\hline

\end{tabular}
\label{Result}
\end{table}

\section{Discussion}

In image-guided interventions, there is an urgent necessity for automatic evaluation of image registration quality. However, the assessment of multi-modal registration results presents notable challenges, primarily stemming from two key factors. First, different contrasts between images need more sophisticated methods to extract useful features for registration error estimation. Second, acquiring accurate ground truths for registration error estimation is often difficult, unlike the tasks of image segmentation or classification. To overcome these challenges, we created a framework by leveraging the Swin UNETR architecture that has demonstrated remarkable effectiveness in various computer vision tasks, owing to its ability to efficiently capture long-range dependencies through the self-attention mechanism of the Swin Transformer and contextual information. Compared to the previous studies, our framework solely worked with real surgical images because the fidelity of simulated ultrasounds remains sub-optimal and can lead to weak performance when applied to real clinical data in testing. Also, in comparison with the tasks of categorical error grading and mean error estimation, our framework can produce a dense error map for image registration, providing the surgeons with more fine-grained information regarding the quality and reliability of the registration results. One of the main limitations of our work arises from the scarcity of patient data, compounded by the variability in settings and properties of US scanners. These could potentially impact the outcome of the proposed technique. In the near future, we plan to explore data-efficient approaches to further refine the methodologies for dense image registration error assessment.

\section{Conclusion}
In this paper, we proposed a robust and efficient method to automatically estimate MRI-iUS registration error in a voxel-by-voxel manner, which is of great importance to ensure the safety and outcomes in ultrasound-guided brain tumor surgery. The results demonstrate that our method can achieve a clinically acceptable outcome and may possess great potential for surgical applications.

\section{Acknowledgement}
We acknowledge the support of the Natural Sciences and
Engineering Research Council of Canada (NSERC) and Fonds de Recherche du Québec Nature et Technologies (FRQNT).

\bibliographystyle{IEEEtran}
\bibliography{ref.bib}

\end{document}